%% file: presence.tex
\documentclass[10pt,twocolumn]{article}
\usepackage{caption}
\usepackage{subcaption}
\usepackage{graphicx,xcolor} 
\usepackage[framemethod=tikz]{mdframed}
\usepackage[font=small,labelfont=bf,tableposition=top]{caption}
\usepackage{flushend}
\usepackage[a4paper, margin=2cm]{geometry}
\usepackage[backend=biber, sorting=nyt, style=authoryear, backref=true]{biblatex}
\usepackage{capt-of, etoolbox}
\usepackage{hyperref}
\usepackage{hypcap}

\title{SaccadeNet: Towards Real-time Saccade Prediction for Virtual Reality Infinite Walking}

\author{
    Yashas Joshi\footnote{yashasjoshi1996@gmail.com}, Charalambos Poullis\footnote{charalambos@poullis.org}\\
    Immersive and Creative Technologies lab\\
    Concordia University\\
}

\date{}

\begin{document}

\makeatletter
\let\@oldmaketitle\@maketitle
\renewcommand{\@maketitle}{\@oldmaketitle
    \includegraphics[width=\textwidth]{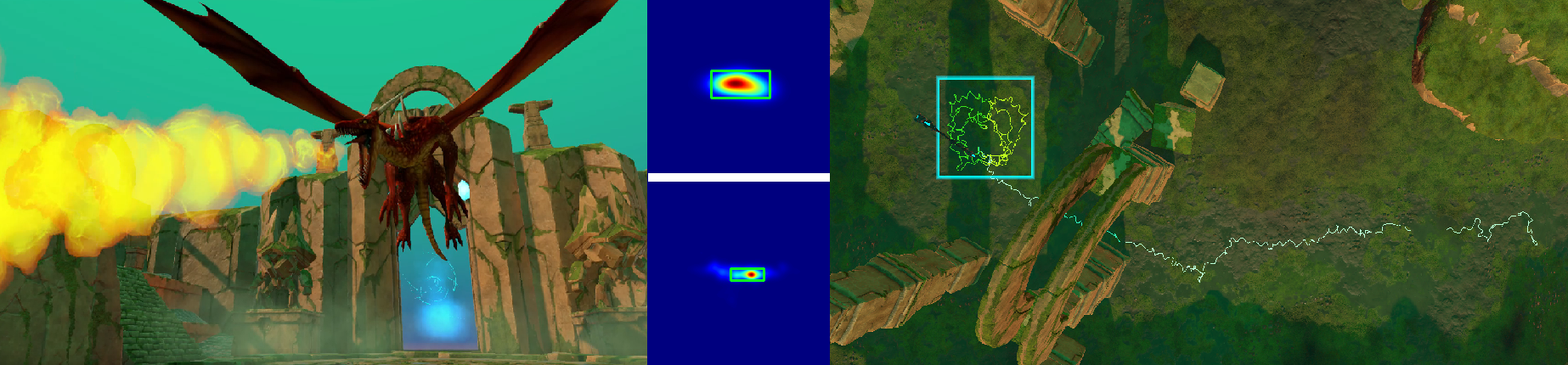}
        \captionof{figure}{Right: A deep-learning powered redirected walking technique that predicts the onset duration of Saccades during head rotations to mask virtual environment manipulations and redirect the users. Mid: average eye gaze positions for two users from our preliminary experiment. Left: Screen capture from a custom-designed VR exploration game that required participants to walk at least 38m straight in the virtual environment from within a 3.5 x 3.5 $m^2$ of physical tracked space.}\label{fig:teaser}}
\makeatother

\maketitle

\begin{abstract}
Modern Redirected Walking (RDW) techniques significantly outperform classical solutions. Nevertheless, they are often limited by their heavy reliance on eye-tracking hardware embedded within the VR headset to reveal redirection opportunities.  
  
We propose a novel RDW technique that leverages the temporary blindness induced due to saccades for redirection. However, unlike the state-of-the-art, our approach does not impose additional eye-tracking hardware requirements. Instead, SaccadeNet, a deep neural network, is trained on head rotation data to predict saccades in real-time during an apparent head rotation. Rigid transformations are then applied to the virtual environment for redirection during the onset duration of these saccades.

We present three user studies. The relationship between head and gaze directions is confirmed in the first user study, followed by the training data collection in our second user study. Then, after some fine-tuning experiments, the performance of our RDW technique is evaluated in a third user study. Finally, we present the results demonstrating the efficacy of our approach. It allowed users to walk up a straight virtual distance of at least 38 meters from within a $3.5 x 3.5m^2$ of the physical tracked space. Moreover, our system unlocks saccadic redirection on widely used consumer-grade hardware without eye-tracking.

\end{abstract}

\input{sec/1_introduction}
\input{sec/2_background}
\input{sec/3_preliminary_experiment}
\input{sec/4_overview}
\input{sec/5_model}
\input{sec/6_experiments}
\input{sec/7_Discussion}
\input{sec/8_conclusion}

\printbibliography

\end{document}

%% file: sec/1_introduction.tex
\section{Introduction}
Over recent years, there have been notable advancements in Virtual Reality (VR) devices due to the advent of GPUs and low-cost displays. As a result, many interaction-based VR applications surged to gain mainstream consumer and industrial attention. People can now explore virtual environments (VEs) from the comfort of their living rooms or offices. However, navigating virtual environments that are spatially larger than the available physical-tracked space (PTS) remains an open research problem. The most common locomotion techniques currently available rely on pointing devices or walking in place. These techniques are unnatural and can negatively impact the sense of presence and immersion since they fail to provide the required inertial force feedback necessary to furnish a sensation of moving into space.

Redirected walking (RDW), circa 2001 \parencite{razzaques2001redirected}, offered a more natural approach to locomotion. In theory, users can navigate an infinite virtual space using RDW while remaining within the boundaries of the available PTS. Since its inception, researchers have proposed many hardware and software-based techniques. 

Hardware-based techniques, e.g. omnidirectional treadmills \parencite{nagamori2005ballArrayTreadmill}, and virtusphere \parencite{fernandes2003cybersphere}, require expensive specialized equipment that allows the user to walk in place. The main drawback of these techniques is that the inertial force feedback provided is not equivalent to natural walking \parencite{christensen2000inertial}, therefore causing vection.

Software-based techniques, although cost-effective, augment the visual content presented to the user. A state-of-the-art technique following a software-based approach is \parencite{sun2018towards}. This technique leverages the natural phenomenon of change blindness induced due to rapid eye movements (saccades) and eye blinks. Observable artifacts are introduced in the original content -as light flashes in 2D or light-orbs in 3D- and serve as visual stimuli to trigger change blindness. Highly efficient eye-trackers embedded in the VR headset continuously track users' eyes in real-time. When a saccade or a blink is detected, the system applies subtle rotations to warp the entire VE while guiding the users away from any potential collision with physical objects, like furniture and walls, detected using a Kinect. Although being apparent, these rotations are imperceptible due to change blindness. However, the main drawback of this technique is that it changes how the user interacts with the virtual environment, deviating from the content creator's intention - for example, navigating around a landmark or playing a game while being distracted by artificially induced visual stimuli.

Following a similar approach, in \parencite{joshi2020inattentional}, an eye-tracking device is embedded in the headset and tracks the user's gaze in real-time while exploiting the psychological phenomenon of inattentional blindness. The system divides the user's virtual field-of-view (FoV) into zones using foveated rendering and continuously updates the peripheral zone, leading to subtle, imperceptible changes due to inattentional blindness. Furthermore, during the temporary blindness triggered due to natural suppressions such as saccades or blinks, the system updates the foveal zone. Hence the entire frame buffer is updated in the direction determined by the RDW algorithm.

The integrated eye trackers required for saccadic redirection and foveated rendering are notably expensive for consumer use. Therefore, this paper investigates the use of deep neural networks for predicting saccades during the users' head rotations. We present experiments that reaffirm the relationship between head and eye rotations and show that users predominantly fixate around the center of the FoV while mainly using head rotations to change their focus. A deep neural network, SaccadeNet, is trained to predict saccades during these head rotations. The VE is adjusted during this phase to redirect the user towards the center of the PTS. Furthermore, as determined by Panero and Zelnik \parencite{panero1979eyeFOV}, the horizontal FoV for a human eye is only $124^{\circ}$. Therefore, to avoid vestibulo-ocular reflexive (VOR) eye movements \parencite{Kothari2020VOR}, we must only predict saccades for the head rotations exceeding $124^{\circ}$ in order to distort the FoV and induce temporary blindness. Thus, we filter the head rotations with velocities exceeding $150^{\circ}/sec$, i.e., more than $124^{\circ}/sec$, and ignore the rest as a limitation of our method. Therefore, this redirection method is effective only during specific tasks that elicit repeated head rotations. On the other hand, this method of predicting saccades can have multiple applications, including foveated rendering and RDW. Finally, we present the evaluation results showing the real-time performance of SaccadeNet. The VE updates were imperceptible to the users, and the system allowed a nearly linear walk in VR while being confined within a $3.5 \times 3.5 m^2$ of PTS. 

%% file: sec/2_background.tex
\section{Background \& Related Work}
Among many interactions, locomotion remains an unsolved impediment to achieving an ideal immersive VR experience. Commonplace techniques like teleportation and flying serve their purpose while relying heavily on external controllers and gamepads. They also often induce simulator sickness and break the presence due to their synthetic feel. On the other hand, an increased sense of presence \parencite{usoh1999flying} and reduced simulator sickness \parencite{laviola2000cybersickness} make natural walking the most favored form of locomotion by the majority of developers. Furthermore, since it is the most natural form of navigation for humans, it also increases the spatial understanding of VEs due to its intuitive nature \parencite{peck2011walkInPlace, ruddle2011cognitiveMapping, ruddle2009cognitiveMapping}.

This section presents a brief overview of the research most relevant to our proposed method. It is categorized in terms of (i) redirected walking, (ii) hardware and software-based solutions, and (iii) natural visual suppressions and state-of-the-art techniques.

\subsection{Redirected walking}

Despite the advantages, locomotion by natural walking poses a significant challenge. Limited availability of the PTS constrains users within a finite boundary, while the virtual space can theoretically be boundless \parencite{poullis2009automatic}. To tackle this dilemma, a solely software-based technique, redirected walking, was introduced circa 2000 \parencite{razzaques2001redirected}.

Visual sense often dominates upon contradicting with vestibular or proprioceptive senses \parencite{VisualVestibularInteractions}. Redirected walking leverages this effect and manipulates the users' virtual FoV such that their physical and virtual motion differ \parencite{15years, Suma2012RDW}. For example, by asking users to walk along predefined curved virtual paths, researchers could inject subtle rotational gains that trick users into taking curved physical paths with a shorter radius \parencite{langbehn2017predefinedCurvedPaths}. These minor discrepancies in the curvatures of physical and virtual paths sufficiently convince users of exploring a comparatively bigger VE than the available PTS \parencite{langbehn2018RDW}.    


The VE rotation is estimated using two main parameters: target directions in VE and PTS. Among many methods of predicting the target direction in the VE, \parencite{zank2015pastWalkingDirection} is one of the approaches that use users' past walking direction, while \parencite{steinicke2008headRotations} uses head rotations, and \parencite{zank2015eyeTracking} the gaze direction. However, algorithms like steer-to-center, steer-to-orbit, and steer-to-multiple-targets \parencite{razzaque2005redirected} are used to determine the target direction in PTS. As the names suggest, steer-to-center algorithms guide users toward the center of the PTS, steer-to-orbit guides them on a fixed orbit around the center of the PTS, and steer-to-multiple-targets guide them towards one of the closest predefined waypoints in the PTS. Experiments performed by Hodgson et al. \parencite{hodgson2013SteerToMultipleCenters} clearly showed a performance advantage of using steer-to-center over other algorithms in vast open virtual spaces. Alignment-based Redirection Controllers (ARC) \parencite{williams2021ARC} is another such algorithm that also avoids obstacles in the physical space by maintaining a similar distance with another virtual object. 

\subsection{Hardware and software-based solutions}
After Razzaque \parencite{razzaques2001redirected} initially paved the way towards solving the enigma of realistic navigation in VR, many researchers followed by either making the technique more efficient or taking a completely different hardware-based approach. 

\noindent
\textbf{Hardware-based approaches:} 
Sliding across a frictionless surface \parencite{iwata1996perambulator} and walking while suspended freely in the air \parencite{benjamin2013suspendedWalking} are prime examples of hardware-based approaches that focus on stationary equipment. In contrast, techniques like omnidirectional treadmill \parencite{souman2010making, darken1997omnitreadmill, iwata1999infiniteFloor, huang2003strollBased, nagamori2005ballArrayTreadmill} and walking in a giant Hamster Ball \parencite{medina2008virtusphere, fernandes2003cybersphere} focus on moving equipment to provide a comparatively realistic experience. Some methods even consolidate external physical props \parencite{cheng2015turkdeck}, or manipulate the entire PTS \parencite{suma2011leveraging,suma2012impossible}.

Although they are plausible solutions to locomotion, they either fail to provide an inertial force-feedback necessary to furnish the sensation of self-motion \parencite{christensen2000inertial}, or have many external dependencies. 

\noindent
\textbf{Software-based approaches:}
These techniques solely rely on digital manipulations for redirection. For example, some scale the user's head rotations and translations in real-time \parencite{razzaques2001redirected, mahdi2017toolkit}, while others rely on producing self-overlapping virtual spaces by partially or fully wrapping the entire VE \parencite{sun2016mapping, dong2017smooth}, or objects in it \parencite{unlimitedCorridor}. 

More recently, techniques like \parencite{langbehn2018blinks} and \parencite{sun2018towards} rely on single subtle VE rotations during a visual suppression, i.e., blink/saccade. Although these subtle techniques are imperceptible and thus preferable in most cases, overt techniques are sometimes favored due to safety or practical limitations. Techniques like freeze-back-up, freeze-turn, and 2:1 turn are some standard approaches proposed by Williams et al. in \parencite{williams2007resetTechniques}. Freeze-back-up allows users to step back upon hitting the PTS boundary with a frozen FoV, freeze-turn allows users to turn by $180^{\circ}$ with a frozen FoV, and 2:1 turn allows users to make half a turn ($180^{\circ}$) in any direction while concurrently making a complete turn ($360^{\circ}$) of the FoV in the opposite direction. In each case, users face the same direction as before in VE with walkable physical space in front. However, the 2:1 turn is exceptionally preferred as it reduces the contradictions between visual and vestibular systems. Moreover, these are generally used in hybrid systems \parencite{joshi2020inattentional} as a reset mechanism when subtle redirection techniques fail. A reset mechanism is mainly employed as a last resort for the safety of the users and equipment. Additionally, open-source platforms like OpenRDW \parencite{li2021openRDW}, and RDW Toolkit \parencite{mahdi2016RDWtoolkit} provide benchmarks and a variety of redirection algorithms for evaluation. 

\subsection{Natural Visual Suppression}
Humans face temporary blindness from time to time \parencite{rensink1997temporaryBlindness, rensink2002temporaryBlindness} due to the actions known as visual suppressions \parencite{volkmann1986suppressions}. Two of the most frequent visual suppressions are blinks and saccades. Blinks are generally essential to maintaining eye functions by spreading tears and removing irritants from their surface. They also modulate cognitive processes such as attentional allocation \parencite{Nakamo2012blink}, and are linked to dopaminergic pathway activation in the human striatum, with clinical conditions in which this pathway activation is increased (e.g., Schizophrenia) or decreased (e.g., Parkinson's) showing concordant changes in blink frequency. On the other hand, Saccades are a necessary eye behavior that allows movement of the relatively small fovea to visual points of interest. They are the ballistic eye movements to change focus from one object to another \parencite{volkmann1986suppressions}. 

With speeds reaching up to 900$^{\circ}$/s \parencite{bahill1975saccadeSpeedRange}, the temporary blindness caused before, during, and after a saccade can last for 20 to 200 ms \parencite{burr1994saccadeTimings}. Tracking some of these fast saccades requires a high refresh rate and high accuracy eye-tracking system. On the contrary, blinks are scarce and more gradual in comparison. The temporary blindness induced due to a blink can typically last for about 100 to 400 ms \parencite{ramot2008blinkDuration}. Users fail to notice any change introduced in the scene during this temporary blindness. This phenomenon is commonly known as change blindness \parencite{regan2000temporaryBlindness}. However, humans make many joint head-eye movements, including not just saccades and blinks, but smooth pursuits and reflexive, compensatory eye movements such as vestibulo-ocular reflexive (VOR) eye movements \parencite{Kothari2020VOR}. Primarily, these oculomotor behaviors do not have similar dynamics. For example, smooth pursuit and VOR movements have lower velocities and accelerations than saccades. The purpose of VOR is to keep an image or object of interest stable on the retina and fovea as the head moves. While visual perception can change during saccades and render temporary blindness, it varies differently for different eye movements.

A method proposed by Langbehn et al. \parencite{langbehn2018blinks} leverages the change blindness induced during naturally occurring blinks to rotate the entire VE. In contrast, Sun et al. \parencite{sun2018towards} leverages the same phenomenon, but with the trigger being saccades instead of blinks. However, the latter relied on simulating artificial saccades by flashing orbs in both image and object space, distracting the users from the task at hand. Following these approaches, Joshi et al. \parencite{joshi2020inattentional} proposed a technique that combines the effects of change blindness with inattentional blindness. The FoV was divided into three zones: peripheral, foveal, and transitional, and rendered using dynamic foveated rendering. Based on their importance, the zones are updated one at a time, slowly replacing the entire frame buffer without noticing. Finally, they update the foveal zone using the temporary blindness caused due to naturally occurring saccades. 

In this paper, we leverage change blindness induced due to saccades and develop a deep learning model for predicting them only during a head rotation. The entire frame buffer is refreshed in a single shot, and hence the user is redirected. 

%% file: sec/3_preliminary_experiment.tex
\section{Head-Eye Relationship}
Humans adjust their gaze continuously by applying simultaneous alterations to our head and eye rotations. The first user study examines this relationship between head and eye directions under the assumption that the users are preoccupied with a cognitive task that elicits repeated head rotations in VR.

\subsection{Application and Procedure}
We developed an immersive VR experience to examine the relationship between head and eye rotations that portrayed an open sky environment to eliminate directional cues. A particular task was designed to ensure numerous simultaneous head rotations during the experiments. Participants were instructed to locate tiny, stationary targets and eliminate them using a semi-automatic firearm. These targets were spawned at a distance of 10 meters from the user. Since saccades tend to be purely eye-driven rather than a joint head-eye movement when the object or image of interest is within 18 deg of fixation point \parencite{stahl1999headEye, Fang2015headEye}, it was also ensured that these targets were separated by at least $20^{\circ}$ from each other. Furthermore, multiple targets in random directions were shown to trigger frequent gaze shifts, resulting in simultaneous head and eye rotations. Upon elimination, the targets were programmed to respawn in another random direction. 

The VR viewport was assumed as a 1x1 grid, and the gaze data was recorded for each participant to generate a heatmap for the entire task duration. The average gaze position was determined by the bounding boxes on each heatmap and plotting a minimum enclosing ellipse. Each test was divided into five repeated steps; a target retrieval task for a one-minute interval, followed by a one-minute break. Therefore, each participant performed the task for five minutes with five breaks. Finally, an elimination score was kept to encourage more target hits in a given time frame. 

\subsection{Participants and Pre-Test Questionnaire}
The study involved 12 participants, two female, with an average age of 24.54 years and a standard deviation of 4.38. In addition, each participant filled out a pre-test and a post-test questionnaire to measure the demographics and simulator sickness levels using Kennedy's simulator sickness questionnaire \parencite{kennedy1993ssq}, respectively. The reported median for their VR device experience was four with normal or corrected-to-normal vision, and their experience using an eye-tracking device was three. This data was gathered using a 5-point Likert scale, with one least and five being most familiar. 

\subsection{Equipment and Safety}

Every experiment during this research was performed on a workstation with an Intel(R) Core(TM) i9 - 9900K CPU @3.60GHz and an NVIDIA RTX 2080 Ti GPU. HTC Vive Pro Eye was used as the primary VR headset with an integrated Tobii Eye Tracker, each operating at a frame rate of 120 Hz with eye-tracking accuracy of $0.5^{\circ}-1.1^{\circ}$. Furthermore, an eye calibration for each participant was performed using the VIVE Pro Eye Setup tool before each experiment in every user study to avoid any noisy gaze estimations. 

Additionally, due to the ongoing coronavirus pandemic, the procedures for all the experiments during this research were scrutinized and approved by the institution's Environmental Health and Safety board (EHS) and the Ethics Research (ER) board. Nevertheless, participants were informed of the risks and their right to quit.

\begin{figure}
\centering
\begin{subfigure}{.23\textwidth}
    \centering
    \includegraphics[width = \textwidth]{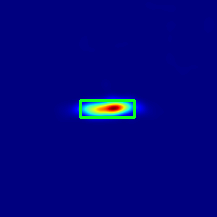}
\end{subfigure}
    \hfill
\begin{subfigure}{.23\textwidth}
    \centering
    \includegraphics[width = \textwidth]{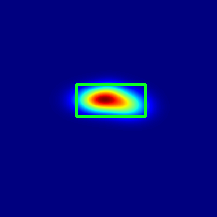}
\end{subfigure}

\caption{Heatmaps and bounding boxes drawn on the FoV of two participants from the first user study. Time spent fixating on the viewport increases from navy blue to dark red.}
\label{fig:heatmaps}
\end{figure}

\subsection{Analysis - Average Foveal Region}
\label{subsec:analysis_first}
Figure \ref{fig:heatmaps} shows the heatmaps for two random participants. Different colors indicate the average time spent by the users fixating on any particular region of the viewport. This time increases as the colors shift from navy blue to dark red. Furthermore, bounding boxes are marked to quantify the region of average gaze. Gartner and Schoenherr's smallest enclosing ellipse algorithm \parencite{Gartner1997smallestenclosing} was then applied to determine the smallest ellipse that encloses every bounding box. This ellipse is given by $\frac{(x-0.491)^2}{0.28^2} + \frac{(y-0.451)^2}{0.144^2} = 1$ with $(0.491, 0.451)$ being its center. Since we initially considered the viewport as a 1 x 1 plane, it is evident that almost all the users mainly focused around its center, i.e. (0.5, 0.5). Therefore, it is safe to assume that the users perform a saccade-like action during an apparent and rapid head rotation. Moreover, since the human's horizontal FoV spans roughly $124^{\circ}$ \parencite{panero1979eyeFOV}, we hypothesize that when the users rotate their heads in VR with a velocity of more than $150^{\circ}/sec$, i.e., more than $124^{\circ}/sec$, the FoV is distorted enough to cause temporary blindness.  



\noindent
\textbf{Simulator Sickness Questionnaire.} 
We used the formulas provided in \parencite{kennedy1993ssq} to calculate the Total Severity (TS) and its corresponding sub-scales such as Nausea (N), Oculomotor (O), and Disorientation (D). A majority ($83.34\%$) of participants reported no significant signs of simulator sickness with an expected highest average score for disorientation ($16.24$) due to the vestibular disturbances caused by repeated head rotations. An overview of these SSQ responses is shown in Table \ref{tab:SSQ1}.

\begin{table}
  \begin{center}
    \caption{Overview of SSQ responses for User study \#1.}
    \label{tab:SSQ1}
    \resizebox{.5\textwidth}{!}{
    \begin{tabular}{|p{70pt} c c c c c|}
      \hline
      \textbf{Scores} & \textbf{Mean} & \textbf{Median} & \textbf{SD} & \textbf{Min} & \textbf{Max}\\
      \hline
      \textbf{Nausea(N)} & 7.95 & 0 & 16.19 & 0 & 57.24\\ 
      \hline
      \textbf{Oculomotor(O)} & 9.475 & 7.58 & 11.71 & 0 & 37.9\\ 
      \hline
      \textbf{Disorient.(D)} & \textbf{16.24} & 0 & 26.43 & 0 & 83.52\\ 
      \hline
      \textbf{Total Score(TS)} & 12.16 & 5.61 & 18.19 & 0 & 63.58\\ 
      \hline
    \end{tabular}
    }
  \end{center}
\end{table}

%% file: sec/4_overview.tex
\section{Technical overview}
To further strengthen the hypothesis formulated in our first study, we developed and trained a deep neural network, SaccadeNet, that predicts saccades solely based on head rotations. Once a saccade is predicted, we adjust the VE according to the Redirection algorithm and redirect the users during the onset duration of these predicted saccades. Indeed, like high-end eye-trackers, SaccadeNet performs in real-time, and our final user study shows that it can redirect users successfully in VR without any noticeable visual artifacts.

%% file: sec/5_model.tex
\section{Model, Data Acquisition \& Learning}
\label{subsec:model_architecture}

This section presents SaccadeNet, a deep neural network that predicts saccades only during tracked head rotations in real-time. It was developed using Pytorch framework and comprised four 1D convolutional layers, a layer to flatten features, and four fully connected layers. The width remains constant (10) throughout the layers as each convolutional layer was specified with a kernel size of 3, and a padding and stride of 1.  Since our data is a time series, the input window size of nine was chosen with consecutive samples selected from the dataset. Therefore, the first layer consisted of nine input channels and 16 output channels. It was followed by (16, 32), (32, 64), and (64, 128) input and output channels (layers) for the second, third, and fourth layers, respectively. The output is flattened to produce a $128x10$ input for the first fully connected layer with 1024 output channels, followed by (1024, 512), (512, 256), (256, 128), and (128, 1) fully connected layers. 

The convolutional layers and the first three fully connected layers employed a Leaky Rectified Linear Unit (Leaky ReLU) activation function in the forward pass to circumvent potential vanishing gradients problems. While the fourth fully connected layer employed ReLU activation to produce entirely positive values for binary classification in the final layer with Sigmoid activation to obtain probabilities between 0 and 1. Adam optimizer was used with a binary cross-entropy (BCE) loss function and 0.001 learning rate for backpropagation. Finally, the model was trained over ten epochs with a batch size of 128.

\begin{table}
  \begin{center}
    \caption{Overview of SSQ responses for User study \#2.}
    \label{tab:SSQ2}
    \resizebox{.5\textwidth}{!}{
    \begin{tabular}{|p{70pt} c c c c c|}
      \hline
      \textbf{Scores} & \textbf{Mean} & \textbf{Median} & \textbf{SD} & \textbf{Min} & \textbf{Max}\\
      \hline
      \textbf{Nausea(N)} & 5.45 & 0 & 7.21 & 0 & 19.08\\ 
      \hline
      \textbf{Oculomotor(O)} & 10.83 & 3.79 & 17.77 & 0 & 60.64\\ 
      \hline
      \textbf{Disorient.(D)} & \textbf{14.91} & 6.96 & 17.66 & 0 & 41.76\\ 
      \hline
  \textbf{Total Score(TS)} & 11.49 & 7.48 & 14.99 & 0 & 48.62\\ 
     \hline
    \end{tabular}
    }
  \end{center}
\end{table}

\subsection{Data Acquisition}
\label{subsec:data_collection}
An application similar to the one in preliminary study was designed to collect data for training SaccadeNet. With the primary task being the same, the tests were divided into three timed trials to reduce the potential of simulator sickness. These trials were timed for five, ten, and fifteen minutes, respectively. A short break followed each trial as per the participants' needs. Furthermore, user interactions in this user study are consistent with the first user study, and participants are under the same experimental constraints. At the end of the experiment, Kennedy's sickness simulator questionnaire \parencite{kennedy1993ssq} was filled to quantify the comfort level during this immersive experience.

A total of 14 participants were recruited for data collection with an average age of 25.86 years and a standard deviation of 4.37. To avoid redundancy in data, each participant recruited for this phase differed from those who participated in the first study. With a normal or corrected-to-normal vision, the reported median for their experiences using a VR device and an eye-tracking device were four and three, respectively on a 5-point Likert scale, with one being least familiar and five being most.

\noindent
\textbf{Simulator Sickness Questionnaire (SSQ)}
Majority (71.43\%) of participants reported no significant signs of simulator sickness with an expected highest average score for disorientation. Table \ref{tab:SSQ2} shows the results of the SSQ responses after data acquisition.

\subsection{Learning \& Inference}
\noindent
\textbf{Input features.}
We trained SaccadeNet on time series data comprised of historical head rotations. Several hand-crafted features were initially computed from head rotations. After confirming the effect of each of those features on saccade detection, the following were eventually filtered for training SaccadeNet. Specifically, we categorized the head's historical fixation direction, its angular velocity, and the acceleration between three successive frames as the most pertinent information for training SaccadeNet. Assuming $f_{t-2}, f_{t-1} and f_t$ denote the last three frames; we filtered the following features:
\begin{enumerate}
  \itemsep-0.2em 
  \item $h_{t-2}$: y component from the head rotation at $f_{t-2}$
  \item $h_{t-1}$: y component from the head rotation at $f_{t-1}$
  \item $h_t$: y component from the head rotation at $f_t$
  \item $\Delta D_y$: Change in direction from $f_{t-1}$ to $f_t$
  \item $V_2$: Angular Velocity from $f_{t-2}$ to $f_{t-1}$
  \item $V_1$: Angular Velocity from $f_{t-1}$ to $f_t$ 
  \item $\Delta V$: Change in Angular Velocity from $V_2$ to $V_1$  
  \item $A_2$: Angular Acceleration from $f_{t-2}$ to $f_{t-1}$
  \item $A_1$: Angular Acceleration from $f_{t-1}$ to $f_t$  
  \item $\Delta A$: Change in Angular Acceleration from $A_2$ to $A_1$  
\end{enumerate}

As humans are least sensitive to horizontal changes, and we only rotate the virtual environment horizontally for redirection, each of these features were measured on the yaw/y/UP axis. Additionally, each feature was recorded independently in world space, i.e. eye-in-world and head-in-world velocities.

These ten features, along with the ground truth of saccadic events, completed a data point saved at each frame. One million nine thousand six hundred and sixty-seven of these data points were saved in total. As determined by Sun et al. \parencite{sun2018towards}, a ballistic eye movement with an angular velocity of more than $180{^\circ}/s$ was classified as a saccade. Therefore, the ground truth was set to 1 in every frame for the onset duration of a saccadic event, and 0 for the rest. Furthermore, since our primary shortcoming is to predict saccades only during an apparent head rotation, any saccadic event during a head rotation velocity of less than $150{^\circ}/sec$ was disregarded, and the ground truth was artificially set to 0 for that frame. Therefore, the model never predicts a saccade when the head rotation velocity is less than $150{^\circ}/sec$. This threshold is from our hypothesis (section \ref{subsec:analysis_first}) and addresses the problem of VOR eye movements. Finally, we normalize the dataset, and using a window size of nine, train SaccadeNet for binary classification with the optimization details explained in Section \ref{subsec:model_architecture}. 

\noindent
\textbf{Training, validation, and testing.} The dataset was divided into three subsets, i.e., training set, validation set, and test set, with an 80:10:10 split ratio. After each epoch, the model was evaluated with the validation set, while the final predictive performance was evaluated using a test set. The average precision during validation was $89.91\%$ and the mean accuracy $93.41\%$., and the precision on the test set was $88.72\%$ and the accuracy $93.51\%$.

\begin{figure}[h]
    \centering
    \includegraphics[width=0.38\textwidth]{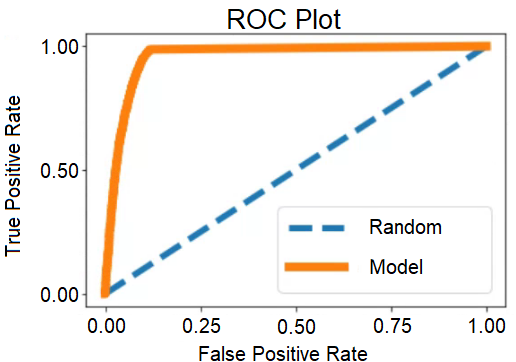}
    \caption{ROC is plotted against FPR on the x-axis and TRP on the y-axis.}
    \label{fig:roc}
\end{figure}

Furthermore, the model was validated using the Area Under the Curve of the Receiver Operating Characteristics curve (AUC ROC). Figure \ref{fig:roc} shows the ROC graph plotted for various thresholds against True Positive Rate (TPR) and False Positive Rate (FPR). The dotted line indicates the worst-case scenario, i.e., completely random predictions (AUC = 0.5), and the orange curve shows the ROC. AUC for the ROC curve was higher than the worst-case and closer to 1, i.e., 0.966. 

%% file: sec/6_experiments.tex
\section{User study}

For the final experiment, we designed an application to evaluate our event-based redirected walking system as a whole. Whenever SaccadeNet predicts a saccade, we adjust the VE and exploit the natural phenomenon of change blindness to hide the redirection, thus ensuring a smooth and distraction-free immersive experience.

\subsection{Application and Procedure}
According to a study published by Simons et al. \parencite{simons2005changeblindness}, change blindness is the inability of observers to notice massive changes in plain sight, which can be attributed to the lack of attention. This phenomenon is commonplace in VR applications where the users are typically engaged in cognitive tasks such as training simulations and games. Therefore, we designed a first-person treasure hunt game on a mysterious island occupied by dragons and swamp crawlers to examine our redirected walking system.  

\noindent
\textbf{Application \& Task.} The main objective in this immersive game is for the user to collect three crystals from the ruins of an ancient abandoned arena. To achieve this, the participants have to walk from their initial spawn position to several predefined locations in VR marked by glowing crystals. At the beginning of the experience, a quick tutorial explains various interactions and the final objective. Upon completing the tutorial, the first crystal and directions to the second crystal are revealed. Each of these crystals unlocks a new magical power. The staff's power to throw lightning bolts is unlocked with the first crystal. \footnote{A gameplay video and several screen captures are included in the supplemental material.}

The effect of change blindness induced due to saccades is further strengthened by introducing a cognitive workload in the form of distractors \parencite{peck2008distractors} when walking towards the second destination. Tiny dragons spawned far away flew with an additional audio cue to a random position in a predefined orbital pattern around the character and were separated by atleast $20^{\circ}$. Figure \ref{fig:teaser} (left) shows a screen capture of a participant's perspective during the experiments. Participants were directed to zap these dragons using their newly gained lightning bolt power triggered by a button on the Vive controller. 

The experience depicted a cognitive task that caused repeated head rotations to increase the probability of predicting a saccade, leading to more redirections. The shortest straight distance to the second destination (38m) was multiple magnitudes larger than the longest possible straight distance in the PTS (4.95m). The cyan-colored box in Figure \ref{fig:teaser} (right) indicates the PTS, and the grey and yellow lines indicate the physical and virtual paths taken by the user during the final experiments, respectively. 

\noindent
\textbf{Redirection algorithm.} A steer-to-center redirection algorithm was implemented with a 2:1 turn reset mechanism in place.

\noindent
\textbf{Procedure.} The study conducted two different experiments, with the proposed redirected walking technique being the independent variable tested. The first experiment involved participants going through the earlier mentioned course of events with an experimental condition, i.e., both redirection and reset mechanism enabled. The second experiment repeated the same course of actions with a controlled condition, i.e., only reset mechanism enabled. At the end of each experiment, relevant information variables such as the number of resets, total distance traveled, and total time taken was gathered for later analysis. Additionally, both the experiments were separated by a short break.


\subsection{Participants}
While designing the experiment, we performed a power analysis with an effect size of 0.7 and determined the sample size to be 20. The power of the experiment was 0.956. We recruited 32 participants, with a mean age of 27.38 years and a standard deviation of 3.74. Twelve of them participated in the fine-tuning experiments and the other 20, 8 female, participated in the final evaluation. The reported medians for their experiences using any VR headset and an eye-tracking device were three and two, respectively, with normal or corrected-to-normal vision. A 5-point Likert scale was utilized to gather this information, with one being the least and five being the most familiar. 

Before the experiment started, participants were informed about the reset mechanism and their objectives in the game. They were also instructed to walk at an average pace and be engaged with the task. Moreover, subjective feedback at the end of the experiment asked the question "Did you notice any visual disparity or shift in the virtual environment?".

\subsection{Analysis}
A detailed analysis of the results obtained from the final study is presented in this section. The first part presents the quantitative results of evaluating the prediction model, SaccadeNet. In the latter part, we show quantitative results of the performance of the proposed redirected walking technique in its entirety.

\subsubsection{SaccadeNet - Quantitative Performance Evaluation}
SaccadeNet performed well within real-time limits with an average accuracy of $94.75\%$, recall of $99.99\%$, and sensitivity of $94.68\%$. The model's F1-score for real-time data was 0.72.
Furthermore, a saccade typically spans over several frames. An analysis of the results showed that the training dataset was highly imbalanced towards the negative training examples, i.e., no saccades, which leads to a higher number of false positives since the true positives are orders of magnitude less. Thus, although the model can predict a saccade correctly, it often mispredicts its duration by a few frames before and after the actual saccade. This causes more false positives. More importantly, it raises the question as to at which point in time one should apply the redirection for it to be imperceptible; at the first positive prediction or after $X$ consecutive positive predictions, and if the latter, what is the best $X$. We perform four fine-tuning experiments involving a total of 12 participants (2+3+3+4) to address this question. These fine-tuning experiments compensate for the $56.52\%$ precision obtained from the final study.

\noindent
\textbf{Fine-tuning.}
In the first experiment, we applied redirection at the first positive prediction. Since the VE was rotated just before the actual saccade began, the two participants (2/2) reported that they noticed the redirection. In the second and third experiments, the VE was rotated only if two and three consecutive frames made positive predictions, respectively. All participants (3/3) in the second and two (2/3) in the third experiment reported noticing the angular shift, while the other participant (1/3) in the third experiment reported a distraction-free experience. In the fourth experiment, redirection was only applied if positive predictions were made over four consecutive frames. All participants (4/4) reported a distraction-free experience and could not perceive any angular shift. 

Given the outcomes of these experiments, we use a window size of four consecutive positive predictions before applying the redirection in the final user study for evaluation.

\begin{figure*}[h]
    \centering
    \begin{subfigure}{0.3\textwidth}
        \centering
        \includegraphics[width=\textwidth]{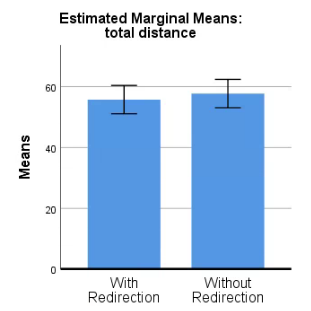}
    \end{subfigure}%
    ~
    \begin{subfigure}{0.3\textwidth}
        \centering
        \includegraphics[width=\textwidth]{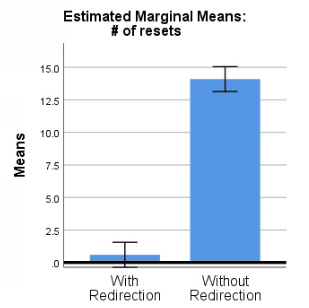}
    \end{subfigure}%
    ~ 
    \begin{subfigure}{0.3\textwidth}
        \centering
        \includegraphics[width=\textwidth]{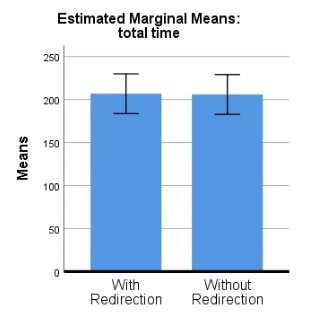}
    \end{subfigure}%
    \caption{Marginal means reported by one-way ANOVA analysis of (a) Distance Travelled, (b) Number of Resets in PTS, and (c) Total Time Taken. Confidence Interval = 95\%}
    \label{fig:ANOVAcharts}
\end{figure*}

\subsubsection{Performance Evaluation of the Proposed Redirected Walking Technique}
We performed a statistical analysis to evaluate the proposed RDW technique. A one-way analysis of variance (ANOVA) was performed between groups, with repeated measures and $\alpha = 0.05$. The analysis statistically differentiated the impacts of using and not using redirection on our dependent variables, such as the number of resets, total distance traveled, and total time taken. The effect sizes for all the variables were determined using partial eta squared $(\eta^{2}_{p})$ values from the analysis. Partial eta squared values reported that our independent variable, i.e., toggling redirection, accounted for almost $86.5\%$ of the observed variance in the number of resets. In comparison, it only contributed about $0.6\%$ and $0.0\%$ to the variances observed in total distance traveled and total time taken, respectively. The analysis also reported a statistically significant difference between the number of resets ($F (1, 62) = 396.094, p < 0.001$), with and without using our proposed redirection technique. However, a statistically insignificant effect of toggling redirection was observed on the total distance travelled ($F (1, 62) = 0.366, p > 0.05$) and time taken ($F (1, 62) = 0.003, p > 0.05$). Means for dependent variables w.r.t. the independent are plotted in figure \ref{fig:ANOVAcharts} with error bars at a confidence interval of $95\%$.

Furthermore, as determined by \parencite{sun2018towards}, the proposed system applied an average absolute gain of $12.59^{\circ}$ per redirection to the VE, at about 0.55 redirections per second. Thus, $1375.1^{\circ}$ of average absolute gain was introduced to each participant’s FoV in total with a standard deviation of $432.131^{\circ}$. Moreover, each participant covered a straight distance of at least 38 meters in the VE by walking within the $3.5 x 3.5 m^2$ PTS. Figure \ref{fig:teaser} (right) shows the path of a participant during the final study. Additionally, as the lower precision was compensated by only redirecting after four consecutive positive predictions, none of the participants noticed any disparity or distraction in the VE. 

\begin{table}[!ht]
  \begin{center}
    \caption{An overview of the SSQ responses.}
    \label{tab:SSQ3}
    \resizebox{.5\textwidth}{!}{
     \begin{tabular}{|p{70pt} c c c c c|}
      \hline
      \textbf{Scores} & \textbf{Mean} & \textbf{Median} & \textbf{SD} & \textbf{Min} & \textbf{Max}\\
      \hline
      \textbf{Nausea(N)} & 10.14 & 9.54 & 12.1 & 0 & 38.16\\ 
      \hline
      \textbf{Oculomotor(O)} & 9.95 & 3.79 & 15.21 & 0 & 60.64\\ 
      \hline
      \textbf{Disorientation(D)} & \textbf{13.05} & 13.92 & 16.18 & 0 & 55.68\\ 
      \hline
  \textbf{Total Score(TS)} & 12.39 & 9.35 & 14.55 & 0 & 56.1\\ 
     \hline
    \end{tabular}
    }
  \end{center}
\end{table}

Alas, most of our participants ($71.87\%$) reported no significant signs of simulator sickness. The mean score for disorientation peaked among the other subscales, similar to our previous studies. However, due to our continuous efforts of reducing this score, we can see it dropping from our first study in Section \ref{subsec:analysis_first} (16.24) to the Data Collection in Section \ref{subsec:data_collection} (14.91), and finally, this final user study (13.05). Table \ref{tab:SSQ3} shows an overview of the SSQ responses for the final user study. Overall, every participant had a smooth experience, with one stating, “I felt like walking straight inside the game but I was actually walking in circles to my surprise. It’s a really good experience and I didn’t notice any distractions.”

%% file: sec/7_Discussion.tex
\section{Discussion}

Our motivation behind this technique is to perform saccade prediction and saccadic redirection on commodity hardware and eliminate the need for specialized hardware, e.g., eye trackers, which is currently the case for state-of-the-art \parencite{sun2018towards}. However, like all RDW techniques, it only works in applications with a moderate cognitive workload. The saccades used to mask the VE rotations are predicted using the head rotation data in real-time. Therefore, the technique is specifically effective when the users are preoccupied with a task that elicits repeated head rotations such as battleground training simulations or game-like scenarios similar to the one used in our final user study. On the contrary, in other situations like indoor explorations and calmer experiences, this technique will fail to perform. Additionally, since the saccades tend to be purely eye-driven rather than a joint head-eye movement when the object of interest is within $18^{\circ}$ of fixation point, the technique will perform best when the targets are separated by $20^{\circ}$. Furthermore, this method of predicting saccades can have multiple applications, including foveated rendering and redirected walking. 

\subsection{Evaluation}
The efficacy of our system is examined in the final user study. We demonstrated that SaccadeNet eliminates the need for high-end eye trackers for redirected walking, and qualitatively and quantitatively evaluated its effectiveness. Results showed that the users could explore long straight virtual distances of at least 38 meters by naturally walking within a room-scale physical tracked space of $3.5x3.5m^2$. Figure \ref{fig:teaser} (right) shows an example of a long, straight, virtual walk with an entirely circular physical path. During data collection, even though artificially saving the ground truth to 0 for the frames when the head rotation velocity is less than $150{^\circ}/sec$ resulted in an imbalanced dataset and the danger of overfitting, the predictions are compensated in our fine-tuning experiments during the final evaluation. Fine-tuning the hyper-parameters led us to only apply redirection if four consecutive positive predictions were made. Resulting in each of the 20 participants completing the task without any distraction, despite the model's 56.52\% precision. Upon asking, "Did you notice any visual disparity or shift in the virtual environment during the entire experience?" in a post-test questionnaire, one participant stated, "I did not notice any visual disparity or shift. The experience was smooth."

\subsection{Limitations}
\subsubsection{Seemingly Forced Head-Rotations}
There could be extreme scenarios with repeated left-right head rotations depending on the location of the enemies. Similarly, there could be no head rotations when the new target location lies within the users' FoV. However, the enemies' placement is random, and we do not force the users to look in any particular direction. Additionally, since the scope of our current technique only includes training simulations and hyperactive game-like experiences, we do not concern with natural head rotations. Nonetheless, we plan to incorporate it in the future.

\subsubsection{Slow User Movement}
Similar to the image and object space subtle gaze directors used by Sun et al. \parencite{sun2018towards}, if the user is walking at a slower pace, there is little any RDW technique can do without introducing distractors and stimulating the required user actions. Moreover, the task designed for the final user study acts as a moderate cognitive workload. It also helps us further strengthen the effect of change blindness by introducing an attention deficit.

\subsubsection{Comparison}
We did not compare directly with other RDW techniques because the comparison would be unfair without embedded eye-trackers. Instead, quantitative analysis with resets-only baseline provides a more accurate measure of effectiveness and allows for more informative comparisons with other techniques.

\subsection{Advantages}

The options for VR headsets with integrated eye-trackers are currently minimal. They are also considerably more expensive than any regular VR headset, limiting their use for mainly research or industrial purposes. Apart from the positional tracking, our technique does not require additional hardware, i.e., eye-trackers or Kinect. It gives us many essential advantages such as computing requirements, accessibility, and hardware cost, while distinguishing our approach to redirection from the other state-of-the-art works reported in the literature review section \parencite{sun2018towards, langbehn2018blinks, langbehn2017predefinedCurvedPaths, joshi2020inattentional}. Moreover, since positional tracking in mobile VR has recently been enabled through SDKs like Google AR Core and ARKit, optimizing the model for cell phone processors can also potentially unlock the possibility of saccade prediction on readily available mobile VR.

%% file: sec/8_conclusion.tex
\section{Conclusion and Future Work}
This work presents a novel event-based redirection technique powered by deep learning. SaccadeNet, a CNN-based model, was trained to predict the change blindness induced due to saccades during a head rotation. Our technique exploits these predicted visual suppressions and repeatedly applies subtle rotations to the VE. These rotations, however subtle, are enough for the users to subconsciously change their physical walking direction while perceiving a straight motion in VR. Three user studies were conducted for (i) reaffirming the relationship between head and eye directions, (ii) training data acquisition for SaccadeNet, and (iii) evaluating the proposed predictor with the redirected walking technique and demonstrating its effectiveness in VR applications with moderate cognitive workload. The studies also confirmed that the proposed method could successfully handle long straight walks, allowing users to freely roam in large scale virtual spaces by walking in applications such as hyper-active immersive games or training simulations.

\vspace{5pt}
\noindent
\textbf{Future Work.} Despite overcoming the eye-tracking hardware requirement for saccadic redirection, many potential avenues for improvement still exist. Firstly, since we only classify saccades during head rotations, we plan to address the exciting problem of predicting saccades in normal viewing conditions. It will also be interesting to explore the correlations of saccades with different scenes and stimulations in the FoV. For example, prediction of saccades while the users attend to various audio-visual stimulations like explosions, flashing lights, or spatial sounds embedded directly into the scene without distracting or changing the 3D content. Additionally, there is also room for improvement in our dataset. We predict saccades solely based on historical head rotation data. Therefore, we plan to add saliency maps and depth maps from the virtual scenes to train robust gaze forecasting models. Furthermore, correlations between a user's torso and gaze have also been examined with some positive outcomes \parencite{sidenmark2019coordination}. We plan to exploit these correlations and train a new model with additional features from hand, foot, and torso movements. Lastly, the scope of redirection in our technique is limited to only VE rotations; we would also like to explore translational gains in the future.